\documentclass[twocolumn,preprint2]{aastex631}

\usepackage{epsfig}
\usepackage{diagbox}
\usepackage{multirow}

\begin{document}

\title{Elemental Abundances from Off-center Carbon Burning in Accreting CO White Dwarfs: Implications for SN 2021yfj-like events}

\author[0000-0002-2452-551X]{Chengyuan Wu}
\affiliation{International Centre of Supernovae (ICESUN), Yunnan Key Laboratory of Supernova Research, Yunnan Observatories, Chinese Academy of Sciences (CAS), Kunming 650216, China}
\email{wuchengyuan@ynao.ac.cn}

\author{Dongdong Liu}
\affiliation{International Centre of Supernovae (ICESUN), Yunnan Key Laboratory of Supernova Research, Yunnan Observatories, Chinese Academy of Sciences (CAS), Kunming 650216, China}

\author{Takashi J. Moriya}
\affiliation{National Astronomical Observatory of Japan, National Institutes of Natural Sciences, 2-21-1 Osawa, Mitaka, Tokyo 181-8588, Japan}
\affiliation{Graduate Institute for Advanced Studies, SOKENDAI, 2-21-1 Osawa, Mitaka, Tokyo 181-8588, Japan}
\affiliation{School of Physics and Astronomy, Monash University, Clayton, VIC 3800, Australia}

\author{Zhengwei Liu}
\affiliation{International Centre of Supernovae (ICESUN), Yunnan Key Laboratory of Supernova Research, Yunnan Observatories, Chinese Academy of Sciences (CAS), Kunming 650216, China}

\author{Heran Xiong}
\affiliation{Research School of Astronomy and Astrophysics, The Australian National University, Canberra, ACT 2611, Australia}

\author{Bo Wang}
\affiliation{International Centre of Supernovae (ICESUN), Yunnan Key Laboratory of Supernova Research, Yunnan Observatories, Chinese Academy of Sciences (CAS), Kunming 650216, China}
\email{wangbo@ynao.ac.cn}

\begin{abstract}

SN 2021yfj is a recently discovered interacting supernova that exhibits narrow emission lines of Si, S, and Ar, indicating the presence of circumstellar material (CSM) enriched with these elements surrounding the progenitor prior to the explosion. The origin of SN 2021yfj-like events remains uncertain. Recent work proposed that the SN 2021yfj-like events may stem from the double WD merger scenario, in which the merger of a Si-rich WD with a more massive ONe WD tidally strips about $0.3{M}_{\odot}$ of Si-rich material to form the CSM. If the merger subsequently triggers a supernova explosion, the interaction between the ejecta and the CSM can reproduce the observed light curve of SN 2021yfj. In this scenario, the progenitor system is a CO WD + He star binary, in which the CO WD accretes He-rich material from the He star. The accumulated material can trigger off-center carbon burning, potentially leading to the formation of a Si-rich WD. However, it remains unclear whether such off-center carbon burning can produce Si, S, and Ar in amounts comparable to those inferred for the CSM of SN 2021yfj. In this work, we simulate the evolution of a CO WD accreting He-rich material using time-dependent mass-accretion rates. Our results show that off-center carbon burning in the accreting CO WD can produce significant amounts of Si and S. We further found that the resulting elemental abundances are strongly affected by the initial carbon abundance of the WD. Based on our results, we suggest that the double WD merger scenario may provide a viable progenitor channel for SN 2021yfj-like events.

\end{abstract}

\keywords{White dwarf stars; Supernovae: general; Supernovae: individual (SN 2021yfj)}

\section{introduction}

Massive stars with initial masses greater than about $8-12{M}_{\odot}$ are believed to explode as core-collapse supernovae (CCSNe). CCSNe are classified into different types according to their spectral properties (e.g., \citealt{1997ARA&A..35..309F}; \citealt{2017hsn..book..195G}; \citealt{2019NatAs...3..717M}). Events that show prominent hydrogen (H) lines in their spectra are classified as Type II SNe, while those that lack H features or only show weak H at early times are referred to as H-poor or stripped-envelope supernovae (SESNe), including Types Ib, Ic, and IIb SNe. Some SNe exhibit signatures of strong interaction between the supernova ejecta and dense circumstellar material (CSM). These interacting SNe are commonly identified by the presence of narrow emission lines in their spectra. For example, events that display bright and narrow Balmer lines are classified as Type IIn SNe, indicating interaction with H-rich CSM. In contrast, events that show narrow helium (He) lines but little or no H lines are classified as Type Ibn SNe, suggesting interaction with a H-poor but He-rich CSM (e.g., \citealt{2007Natur.447..829P}). Some SESNe even show narrow emission lines of carbon (C), oxygen (O), and neon (Ne) in their spectra. These events are referred to as Type Icn SNe (e.g., \citealt{2021arXiv210807278F}; \citealt{2022ApJ...927..180P}; \citealt{2022Natur.601..201G}; \citealt{2022ApJ...938...73P}; \citealt{2023A&A...673A..27N}; \citealt{2023MNRAS.523.2530D}), implying interaction between the SN ejecta and CO-rich CSM.

Recently, \cite{2025Natur.644..634S} reported the discovery of SN 2021yfj, which exhibits narrow emission lines of silicon (Si), sulfur (S) and argon (Ar) in its spectra. These features indicate interaction between the supernova ejecta and a dense CSM enriched in intermediate-mass elements such as Si and S. Owing to its unusual spectral properties and CSM composition, this event has been proposed as a new class of interacting supernova, referred to as Type Ien SN. Observations on the light curve of SN 2021yfj suggests a rapid rise time of $2.3$ days, a peak luminosity of $(3-5)\times{10}^{43}\,{\rm {erg}}\,{\rm {s}^{-1}}$ and a radiated energy of $(0.5-1.0)\times{10}^{50}\,{\rm {erg}}$. By modeling the early-time spectra of SN 2021yfj, Schulze et al. (2025) inferred that the CSM has mass fractions of approximately $78.6\%$ (O), $10\%$ (Ne), $5\%$ (Si), $3\%$ (S), $1\%$ (Ar), $1\%$ (Mg), and $0.1\%$ (Ca), while iron-group elements (Fe, Co and Ni) are consistent with solar abundances.

Owing to the unusual spectral properties of SN 2021yfj, its progenitor remains uncertain, particularly when the He lines is also found in the spectra. Several progenitor channels have been proposed (e.g., \citealt{2025Natur.644..634S}), although each faces certain challenges. (1) One possibility is that SN 2021yfj originates from a pulsational pair-instability supernova (e.g., \citealt{2017ApJ...836..244W}). In this scenario, a massive star loses its H envelope through stellar winds, eruptions, or binary interaction, forming a massive He star. Such a He star may undergo multiple pulsational pair-instability episodes during advanced burning stages, ejecting shells prior to the core collapse. The interaction between these shells can reproduce several observed properties of SN 2021yfj, including the wind velocity, the presence of Si, S, and Ar, as well as its rise time and luminosity. However, this scenario has difficulty explaining the presence of He in the spectra. (2) Another possibility involves a relatively low-mass He star with a mass of $2.0-2.6\,{M}_{\odot}$. In this case, degenerate Si flashes may drive strong mass loss, ejecting Si-rich material prior to a final iron-core collapse (e.g., \citealt{2015ApJ...810...34W}). The subsequent interaction between the SN ejecta and the surrounding material could account for some observational features of SN 2021yfj. However, this scenario suffers from significant uncertainties in the energy release and timing of the Si flashes, and it is considered more relevant for Type Ibn SNe. (3) A third possibility is related to jet-driven explosions in massive stars. In this scenario, SN 2021yfj would need to be observed along a preferred viewing angle close to the jet axis. However, such explosions are often expected to produce $\gamma$-ray flashes, which has not been detected for SN 2021yfj. In addition, jet-driven outflows may dredge up inner material rich in C, O, and Ne, whereas strong emission lines of these elements are not observed in SN 2021yfj. (4) Finally, the merger of two compact objects has also been proposed. In this scenario, Si and S may be synthesized on the surface of a WD or possibly a neutron star. However, simulations of He burning on the surface of CO WDs found that the production of Si and S is subdominant compared to the elements such as Ca, Mg, and iron-group elements, making it difficult to account for the observed CSM composition.

Recently, \cite{2026arXiv260307064M} proposed that the observed properties of SN 2021yfj can be reproduced by the interaction between supernova ejecta and a dense Si/S-rich CSM with a mass of $\sim0.3{M}_{\odot}$. Their model adopts an explosion with a kinetic energy of $\sim4.0\times{10}^{50}\,{\rm {erg}}$ and ejects $\sim0.3{M}_{\odot}$ of material, pointing to a double WD system as a viable progenitor channel. In this scenario, the progenitor system is a CO WD+He star binary, consisting of a $1.1{M}_{\odot}$ CO WD and a $2.3{M}_{\odot}$ He star with an initial orbital period of $0.25$ days. As the He star evolves into the giant phase, it fills its Roche lobe and initiates stable mass transfer. The accumulated He triggeres off-center carbon burning in the accreting CO WD (e.g., \citealt{2016ApJ...821...28B}; \citealt{2017MNRAS.472.1593W}; \citealt{2021ApJ...922..241W}), leading to the formation of a Si-rich envelope (e.g., \citealt{2019MNRAS.486.2977W}; \citealt{2020MNRAS.495.1445W}). After the mass-transfer phase, the system evolves into a double WD binary consisting of a $1.28{M}_{\odot}$ Si-rich WD and a $1.33{M}_{\odot}$ ONe WD with an orbital period of $0.18$ days. Gravitational wave radiation then drives the merger of the two WDs. During the merger process, $\sim0.3{M}_{\odot}$ of material from the outer layers of the less massive Si-rich WD is tidally stripped and forms the CSM, while the remaining material is accreted onto the more massive ONe WD. The subsequent merger-triggered thermonuclear explosion ejects $\sim0.3{M}_{\odot}$ of material and produces observational properties similar to those of SN 2021yfj, leaving behind a bound remnant with a mass of $\sim2.0{M}_{\odot}$. The surviving remnant may continue to evolve and potentially undergo electron-capture collapse on a timescale of $\sim100-1000$ years after the merger (e.g., \citealt{2023ApJ...944L..54W}). Meanwhile, according to \cite{2026arXiv260307064M}, the delay time of such a progenitor system is approximately $\sim3.3\,{\rm {Gyr}}$. \cite{2025Natur.644..634S} estimated the age of the host galaxy of SN 2021yfj is about ${\rm 4}^{\rm {+4}}_{\rm {-2}}\,{\rm {Gyr}}$. Therefore, the delay time of this double WD merger scenario is broadly consistent with the age of the host galaxy, supporting the possibility that SN 2021yfj originated from such a progenitor system.

In the double WD merger scenario described above, the formation of the Si-rich WD relies on off-center carbon burning in a He-accreting CO WD. In \cite{2026arXiv260307064M}, the evolution of the accreting WD was modeled assuming a constant mass-accretion rate of $5.0\times{10}^{-6}\,{M}_{\odot}\,{\rm {yr}^{-1}}$. However, in realistic binary evolution, the mass-transfer rate and hence the accretion rate onto the CO WD varies significantly with time. Previous studies have shown that the nucleosynthesis during off-center carbon burning is highly sensitive to the mass-accretion rate (e.g., \citealt{2019MNRAS.486.2977W}; \citealt{2020MNRAS.495.1445W}). Therefore, adopting a constant accretion rate may lead to uncertainties in the predicted elemental abundances, which introduces uncertainty in assessing whether the double WD merger scenario can account for the observed properties of SN 2021yfj.

In this work, we plan to simulate the evolution of He-accreting CO WDs by using time-dependent mass-accretion rates derived from binary evolution calculations. Our goal is to assess whether the evolution of He-accreting CO WD is consistent with that predicted by binary evolution, and whether the resulting nucleosynthetic yields from off-center carbon burning can reproduce the elemental abundances inferred for SN 2021yfj.

The article is organized as follows. In Sect.\,2, we describe the basic assumptions and numerical setup for modeling He-accreting CO WD. In Sect.\,3, we present the main results of the CO WD evolution. We discuss the uncertainties of the progenitor model in Sect.\,4 and summary are given in Sect.\,5.

\section{Methods}

In the double WD merger scenario for producing SN 2021yfj, the Si-rich WD is formed through He accretion onto a CO WD. During the mass-transfer phase, He burning on the WD surface is strongly regulated by the mass-accretion rate (\citealt{2014MNRAS.445.3239P}). When the accretion rate is lower than the lower limit for steady He burning ($\dot{M}_{\rm {acc}}<\dot{M}_{\rm {st}}$), He burning becomes unstable and leads to recurrent He-shell flashes, which expel part of the accreted material. In this case, the WD mass growth rate can be written as $\dot{M}_{\rm {WD}}={\eta}_{\rm {He}}\dot{M}_{\rm {acc}}$, where ${\eta}_{\rm {He}}$ is the mass accumulation efficiency during He-shell flashes (e.g., \citealt{2004ApJ...613L.129K}; \citealt{2017A&A...604A..31W}). When the accretion rate exceeds the upper boundary for steady helium burning ($\dot{M}_{\rm {acc}}>\dot{M}_{\rm {cr}}$), the WD cannot process all the accreted helium via nuclear burning, leading to envelope expansion toward a He-giant configuration. The critical accretion rate $\dot{M}_{\rm {cr}}$ can be approximated by
\begin{equation}
    \dot{M}_{\rm {cr}}=7.2\times{10}^{-6}(\frac{{M}_{\rm {WD}}}{{M}_{\odot}}-0.6){M}_{\odot}/{\rm {yr}},
\end{equation}
which is valid for CO WDs with masses in the range $0.75-1.38\,{M}_{\odot}$ (e.g., \citealt{1982ApJ...253..798N}). Under the optically thick wind assumption, He-rich material is assumed to burn steadily at a rate of $\dot{M}_{\rm {cr}}$, while the excess material is lost via winds at a rate of $\dot{M}_{\rm {acc}}-\dot{M}_{\rm {cr}}$ (e.g., \citealt{1996ApJ...470L..97H}; \citealt{2009MNRAS.395..847W}; \citealt{2018MNRAS.473.5352L}). Therefore, stable He burning occurs only within the range $\dot{M}_{\rm {st}}<\dot{M}_{\rm {acc}}<\dot{M}_{\rm {cr}}$. Adopting this framework, \cite{2026arXiv260307064M} performed binary evolution calculations for a CO WD+He star system consisting of a $1.1{M}_{\odot}$ CO WD and a $2.3{M}_{\odot}$ He star with an initial orbital period of ${P}_{\rm {orb}}^{\rm i}=0.25\,{\rm d}$, where the accreting WD was treated as a point mass. Their results show that the mass-transfer phase lasts for $\sim4.3\times{10}^{4}\,{\rm {yr}}$, during which the CO WD grows to $1.28{M}_{\odot}$, while the He star evolves into a $1.33{M}_{\odot}$ ONe WD. The corresponding mass-transfer rate is shown as the black solid line in Fig.\,1. In this work, we adopt the time-dependent mass-accretion rate derived from their binary evolution calculations to model the detailed evolution of the accreting CO WD.

We use the stellar evolution code ${\tt {MESA}}$ version $10398$ (e.g., \citealt{2011ApJS..192....3P}; \citealt{2013ApJS..208....4P}; \citealt{2015ApJS..220...15P}; \citealt{2018ApJS..234...34P}) to simulate the evolution of a He-accretion CO WD. To construct the initial WD model, we first evolve a $1.1{M}_{\odot}$ pre-He-MS star (a phase analogous to the pre-MS phase) with metallicity $Z=0.02$ until its central density reaches ${\rho}_{\rm c}=5.0\,{\rm g}\,{\rm {cm}^{-3}}$. We then relax the chemical composition to a uniform mixture of $^{\rm {12}}{\rm C}$ ($40\%$) and $^{\rm {16}}{\rm O}$ ($60\%$) throughout the star, and allow the model to cool until the WD luminosity decreases to ${L}_{\rm {WD}}=10{L}_{\odot}$ (fiducial model). Nuclear reactions are turned off during this relaxation procedure. In addition to this fiducial model, we construct two additional models with initial compositions of $^{\rm {12}}{\rm C}$ ($30\%$), $^{\rm {16}}{\rm O}$ ($70\%$), and $^{\rm {12}}{\rm C}$ ($50\%$), $^{\rm {16}}{\rm O}$ ($50\%$), respectively, to investigate the impact of the initial carbon abundance on off-center carbon burning.

Starting from this initial model, we accrete He-rich material with a composition of $98\%$ He and $Z=0.02$. The mass-accretion rate is taken from the time-dependent mass-transfer rate obtained in the binary evolution calculations (Fig.\,1). At the onset of accretion, the first He-shell flash is sufficiently strong to drive substantial mass loss. Therefore, we adopt the super-Eddington wind prescription to remove the excess material. After this initial phase, the accretion rate remains above the upper boundary for steady He burning ($\dot{M}_{\rm {acc}}>\dot{M}_{\rm {cr}}$), and we switch to the optically thick wind scheme. To model the optically thick wind, we follow the prescription of \cite{2019ApJ...878..100W}. When $\dot{M}_{\rm {acc}}>\dot{M}_{\rm {cr}}$, the effective mass growth rate of the WD is given by $\dot{M}_{\rm {WD}}=(1-{\beta})\dot{M}_{\rm {acc}}$, where $\beta$ represents the fraction of the mass-transfer rate lost from the system. The parameter $\beta$ depends on the WD radius: it is set to zero when ${R}_{\rm {WD}}<{R}_{\rm {min}}$ and increases smoothly to unity as ${R}_{\rm {WD}}$ approaches ${R}_{\rm {max}}$. Here we adopt ${R}_{\rm {min}}=2{R}_{\rm {WD}}^{\rm i}$ and ${R}_{\rm {max}}=10{R}_{\rm {WD}}^{\rm i}$, where ${R}_{\rm {WD}}^{\rm i}=0.008{R}_{\odot}$ is the radius of the initial cold WD model, corresponding to ${R}_{\rm {min}}=0.016{R}_{\odot}$ and ${R}_{\rm {max}}=0.08{R}_{\odot}$. The choice of ${R}_{\rm {min}}$ and ${R}_{\rm {max}}$ has a negligible effect on the WD mass growth rate. As noted by \cite{2016ApJ...821...28B}, the WD expands very rapidly once the accretion rate exceeds the stability limit, making the exact choice of transition radius unimportant. Consequently, the resulting mass accumulation efficiency is comparable to that adopted in \cite{2026arXiv260307064M}. This supports the use of the mass-accretion rate derived from binary evolution, where the WD is treated as a point mass, as a reasonable approximation in our calculations.

He burning on the WD surface converts the accreted He into C and O. As the WD mass increases, off-center carbon burning is subsequently triggered in the newly formed CO layer. The off-center C flame propagates inward in mass coordinate, and we terminate the calculation when the flame reaches ${M}_{\rm r}\approx0.3{M}_{\odot}$, by which point the nucleosynthesis in the outer layers has been largely established. To investigate the nucleosynthesis during this phase, we adopt the ``approx21.net'' nuclear reaction network in ${\tt {MESA}}$ default. This network includes the key isotopes involved in He and C burning, such as $^{\rm 4}{\rm {He}}$, $^{\rm {12}}{\rm C}$, $^{\rm {16}}{\rm O}$, $^{\rm {20}}{\rm {Ne}}$, $^{\rm {24}}{\rm {Mg}}$, $^{\rm {28}}{\rm {Si}}$, $^{\rm {32}}{\rm S}$, $^{\rm {36}}{\rm {Ar}}$, $^{\rm {40}}{\rm {Ca}}$, as well as representative iron-group elements. It is therefore sufficient to capture the production of intermediate-mass elements relevant for comparison with the abundances inferred for SN 2021yfj. Previous works have shown that the choice of nuclear reaction network may affect the detailed energy generation and abundance patterns, especially during advanced burning stages (e.g., \citealt{2016ApJS..227...22F}). To investigate the influence of the adopted reaction network on the abundance distribution in our models, we also recalculated the fiducial model using a 57-isotope network consisting of ${\rm {n}}$, $^{\rm {{1}-{2}}}{\rm H}$, $^{\rm {{3}-{4}}}{\rm {He}}$, $^{\rm 7}{\rm {Li}}$, $^{\rm {{7},{9}-{10}}}{\rm {Be}}$, $^{\rm 8}{\rm B}$, $^{\rm {{12}-{13}}}{\rm C}$, $^{\rm {{13}-{15}}}{\rm N}$, $^{\rm {{14}-{18}}}{\rm O}$, $^{\rm {{17}-{19}}}{\rm F}$, $^{\rm {{18}-{22}}}{\rm {Ne}}$, $^{\rm {{23}-{24}}}{\rm {Na}}$, $^{\rm {{23}-{25}}}{\rm {Mg}}$, $^{\rm {27}}{\rm {Al}}$, $^{\rm {{27}-{28}}}{\rm {Si}}$, $^{\rm {{30}-{31}}}{\rm P}$, $^{\rm {{31}-{32}}}{\rm S}$, $^{\rm {35}}{\rm {Cl}}$, $^{\rm {36}}{\rm {Ar}}$, $^{\rm {39}}{\rm K}$, and $^{\rm {40}}{\rm {Ca}}$\footnote{The {\tt {MESA}} inlists and initial models used in this work are publicly available \dataset[doi:10.5281/zenodo.20319707]{https://zenodo.org/records/20319708}}.

\section{Results}

In the binary evolution model of Moriya et al. (2026), a $1.1{M}_{\odot}$ point-mass WD accretes from a $2.3{M}_{\odot}$ He star. As the donor evolves into the giant phase and fills its Roche lobe, the mass-transfer rate rises rapidly. Within $\sim350$ years, the transfer rate exceeds the upper limit for stable He burning, driving the accreting WD into the optically thick wind regime. The mass-transfer rate continues to increase and reaches a peak value of ${\log}({\dot{M}_{\rm {tran}}}/{M}_{\odot}\,{\rm {yr}^{-1}})=-4.234$ at ${\rm t}\approx7700$ years. It then gradually declines to ${\log}({\dot{M}_{\rm {tran}}}/{M}_{\odot}\,{\rm {yr}^{-1}})\approx-4.8$. The entire mass-transfer phase lasts for $\sim4.3\times{10}^{4}$ years, after which the He star evolves into an ONe WD. The corresponding mass-transfer history is shown in Fig.\,1.

Using the time-dependent mass-accretion rate derived from the binary evolution, we simulate the evolution of the accreting CO WD. The evolutionary track in the Hertzsprung-Russell diagram (HRD) is shown in panel (a) of Fig.\,2. At the onset of accretion, He-rich material accumulates on the WD surface. As He burning is ignited, the envelope expands, leading to a decrease in the effective temperature and luminosity. Meanwhile, energy generated in the burning layer is transported outward by convection, causing a rapid increase in luminosity. Once the luminosity exceeds the Eddington limit ($L_{\rm Edd}$), a super-Eddington wind is triggered, which removes part of the accreted material during the first He-shell flash. The WD expands to a maximum radius of $\sim0.5{R}_{\odot}$ in this phase. After the initial flash, the WD contracts and continues accreting under the optically thick wind regime. During most of the subsequent evolution, the WD radius remains within the range ${R}_{\rm {min}}-{R}_{\rm {max}}$, indicating that the wind-regulated accretion proceeds in a quasi-steady manner. At ${\rm t}\sim3.0\times{10}^{4}$ years, when the WD mass reaches $\sim1.213{M}_{\odot}$, He burning becomes unstable, resulting in repeated expansions and contractions of the WD radius (panel b of Fig.\,2).

As He burning proceeds, the accreted He-rich material is converted into a CO layer. With continued accretion, the temperature in this layer gradually increases. After $\sim3.8\times{10}^{4}$ yr of accretion, when the WD mass reaches $\sim1.246{M}_{\odot}$, off-center carbon ignition occurs at a mass coordinate of ${M}_{\rm r}\approx1.2{M}_{\odot}$ (red triangles in Fig.\,2). Following the carbon ignition, an inwardly propagating carbon flame is established. At the early stage of the flame propagation, carbon burning primarily produces O, Ne, and Mg, while the abundance of Si remains relatively low ($\sim2\%$; panel a2 of Fig.\,3). As the flame propagates inward, the temperature at the burning front increases, reaching ${\rm {log}}({T}_{\rm {max}}/{\rm K})>9.2$ (panel c of Fig.\,2). Under these conditions, the production of intermediate-mass elements becomes more efficient, leading to enhanced abundances of Si-group elements such as $^{\rm {28}}{\rm {Si}}$ and $^{\rm {32}}{\rm S}$ (panel b2 of Fig.\,3). At such high temperatures, the off-center burning flame becomes highly unstable, leading to fluctuations in ${T}_{\rm {max}}$ (e.g., \citealt{2016MNRAS.463.3461S}; \citealt{2019MNRAS.483..263W}; \citealt{2020MNRAS.495.1445W}). The carbon-burning phase lasts for $\sim4\times{10}^{3}$ years, until the flame reaches ${M}_{\rm r}\approx0.3{M}_{\odot}$, where we terminate the calculation. Owing to the rapid inward propagation of the carbon flame, the total mass growth of the WD during this phase is negligible ($\sim8\times{10}^{-4}\,{M}_{\odot}$).

The evolution of the WD mass under the optically thick wind prescription is shown as the blue solid line in Fig.\,1. For comparison, the mass growth predicted in the binary evolution model, where the WD is treated as a point mass using the prescription of \cite{1982ApJ...253..798N}, is shown as the dashed–dotted line. The two results are broadly consistent, with only a small difference in the final WD mass. This suggests that the WD mass-growth rate in our simulations closely follows that obtained from the binary evolution calculation, indicating that our models capture the carbon flame evolution of the accreting WD under realistic binary mass-transfer conditions. Our results further indicate that such binary evolution can lead to the formation of Si-rich white dwarfs. After the carbon flame reaches the center, the Si-rich WD can continue to accrete He-rich material from its companion. According to the binary evolution model of \cite{2026arXiv260307064M}, the He-star companion still has a mass of approximately $1.4{M}_{\odot}$ at this stage. The subsequent mass-transfer phase lasts for another $\sim4000$ years, during which the Si-rich WD accretes an additional $\sim0.035{M}_{\odot}$ of He-rich material before the companion evolves into an ONeMg WD. During this phase, stable He burning forms a CO-rich layer on the surface of the massive WD. However, previous studies have shown that carbon burning in the newly formed CO layer can further process carbon into intermediate-mass elements, including Si-rich material (e.g., \citealt{2024ApJ...975..186Z}). Therefore, although the post-He-accretion phase may somewhat affect the inferred CSM abundance in our model, the overall influence is expected to be limited because the accumulated shell mass is relatively small and the newly formed layer itself also contains intermediate-mass elements.

In our fiducial model, the initial composition of the CO WD is $^{\rm {12}}{\rm C}=40\%$ and $^{\rm {16}}{\rm O}=60\%$. Since the off-center carbon flame propagates through the CO core, the initial C/O ratio is expected to influence the nucleosynthetic yields. To quantify this effect, we construct additional CO WD models with different initial compositions, namely $^{\rm {12}}{\rm C}=30\%$, $^{\rm {16}}{\rm O}=70\%$, and $^{\rm {12}}{\rm C}=50\%$, $^{\rm {16}}{\rm O}=50\%$. For all models, we adopt the same mass-accretion history and wind mass-loss prescription as in the fiducial case. We find that the location and timing of off-center carbon ignition are nearly unchanged, and the flame propagates inward in a similar manner. Since the nuclear energy generation rate of carbon burning scales approximately as ${\epsilon}_{\rm {nuc}}\propto{\rho}{\chi}^{\rm 2}(^{\rm {12}}{\rm C}){T}^{\rm n}$, where ${\rho}$ is the density, ${\chi}$ is the carbon mass fraction, ${T}$ is the temperature of the flame, ${\rm n}\approx40$ for carbon burning at ${\rm {log}}T=8.8-8.9$ (e.g., \citealt{2013ApJ...772...37D}), models with higher carbon abundances tend to develop slightly higher off-center flame temperatures. However, the overall differences in flame temperature remain modest. Despite the similarity in the overall evolution, the nucleosynthetic yields show a clear dependence on the initial carbon abundance. Models with higher initial carbon fractions produce significantly larger amounts of Si-group elements, including $^{\rm {28}}{\rm {Si}}$, $^{\rm {32}}{\rm S}$, and $^{\rm {36}}{\rm {Ar}}$. This trend can be understood as a consequence of more efficient carbon burning, which drives the nuclear processing toward heavier intermediate-mass elements at higher temperatures.

In the double WD merger scenario proposed for SN 2021yfj, approximately $0.3{M}_{\odot}$ of Si-rich material is expected to be tidally stripped during the merger to form the CSM, while a comparable amount of material is ejected during the explosion (e.g., \citealt{2026arXiv260307064M}). Under this framework, the outer $\sim0.3{M}_{\odot}$ of the accreting CO WD provides a reasonable proxy for the composition of the CSM. We therefore calculate the average elemental abundances in the outer $0.3{M}_{\odot}$ for our models and summarize the results in Table\,1, together with the abundances inferred from spectral modeling. Overall, the different initial C/O ratios have little impact on the abundances of $^{\rm {20}}{\rm {Ne}}$ and $^{\rm {40}}{\rm {Ca}}$. The predicted Ne abundance is broadly consistent with the value inferred from the observation, whereas Ca is underproduced by about an order of magnitude. However, no clear calcium features have been identified in the observed spectra of SN 2021yfj. Although a calcium abundance of $0.1\%$ was adopted in the spectral modeling of \cite{2025Natur.644..634S}, it does not produce prominent spectral features. Therefore, the relatively low calcium abundance predicted by our models remains compatible with the current observations. The abundances of Si-group elements show a stronger dependence on the initial carbon fraction. Increasing the initial carbon abundance enhances the production of Si, S, and Ar, while reducing the abundances of O and Mg. Among our models, the case with higher initial carbon abundance produces S abundance closer to those inferred from the spectral modeling of SN 2021yfj. However, this model also tend to produces larger Si abundance, which may result in somewhat stronger Si features than observed. Nevertheless, as discussed by \cite{2025Natur.644..634S}, increasing the Si and S mass fractions to $30-50\%$ produces stronger Si and S features, although the disagreement with the observations remains moderate. On the other hand, reducing the abundances to below $\sim1\%$ leads to weaker features that are also somewhat inconsistent with the observations. This suggests that the Si and S abundances inferred from the spectra are constrained only within a relatively broad range. Therefore, the elemental abundances predicted by our models remain broadly compatible with the current observations. Moreover, Ar is systematically underproduced, while Mg is overproduced compared to the inferred CSM composition. Although none of our models perfectly reproduces the observed abundance pattern, they demonstrate that off-center carbon burning in accreting CO WDs can naturally produce the key intermediate-mass elements (Si, S, and Ar) required for the CSM. In addition, helium transferred from the companion can remain in the outermost layers, potentially explaining the He lines observed in the spectra of SN 2021yfj. These properties support the double WD merger scenario as a plausible progenitor channel for SN 2021yfj-like events.

Previous studies have shown that the choice of nuclear reaction network can affect the detailed energy generation and abundance distributions during neon, oxygen, and silicon burning stages (e.g., \citealt{2016ApJS..227...22F}). We therefore recalculated the fiducial model using a 57-isotope network. The average elemental abundances in the outer $0.3{M}_{\odot}$ of the recalculated model are compared with those of the fiducial model in Table\,2. In the recalculated model, the calcium abundance is approximately one order of magnitude higher than that in the fiducial model, but its mass fraction still remains below $0.1\%$. The abundances of the other major elements show only minor differences. Overall, the abundance distribution remains broadly unchanged, suggesting that the choice of nuclear reaction network does not significantly affect the main conclusions of the present work.

\begin{figure*}
\begin{center}
\epsfig{file=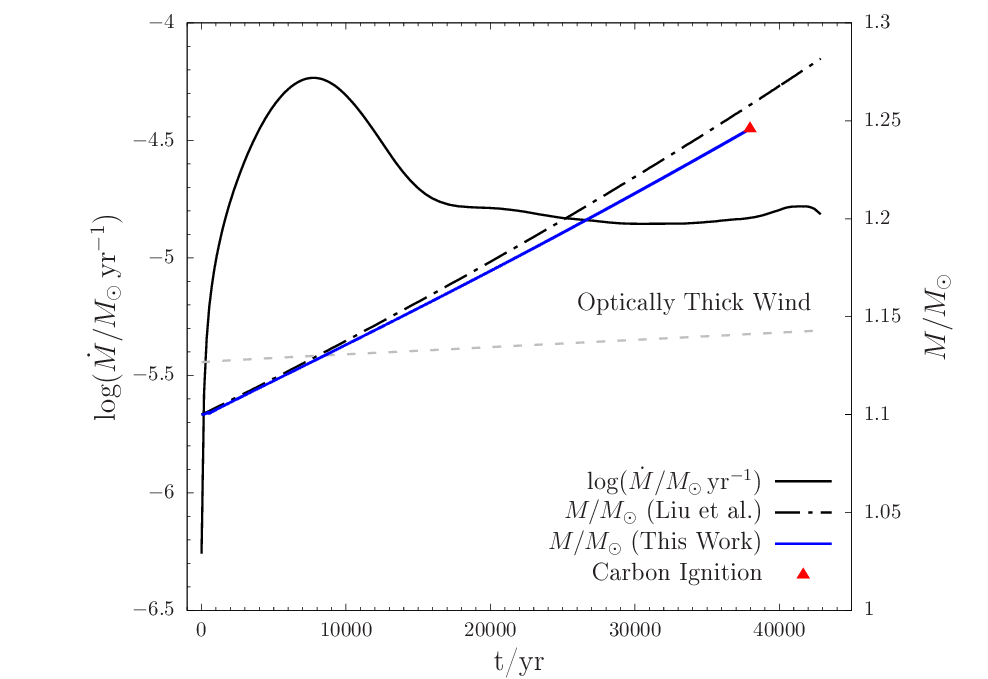,angle=0,width=16.2cm}
 \caption{Mass-transfer and mass-growth rates for the $1.1{M}_{\odot}$ accreting CO WD with an initial composition of $^{\rm {12}}{\rm C}=40\%$ and $^{\rm {16}}{\rm O}=60\%$. The black solid line shows the mass-transfer rate from the binary evolution of \cite{2026arXiv260307064M}, while the black dashed-dotted line represents the corresponding WD mass-growth rate in the binary model. The blue solid line shows the mass-growth rate obtained from our simulations. The grey dashed line indicates the upper limit for stable He burning ($\dot{M}_{\rm {cr}}$), above which the optically thick wind is assumed. The red triangle marks the onset of off-center carbon ignition.}
  \end{center}
\end{figure*}

\begin{figure*}
\begin{center}
\epsfig{file=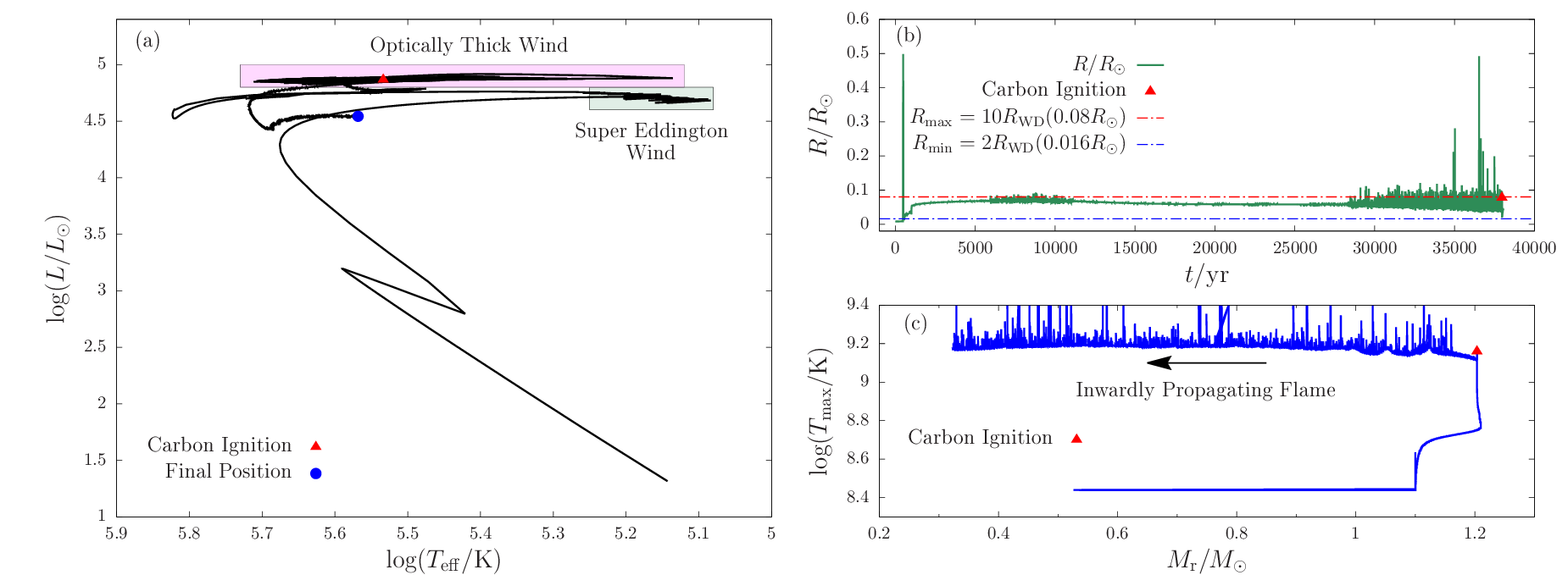,angle=0,width=17.5cm}
 \caption{Evolution of the accreting $1.1{M}_{\odot}$ CO WD. Panel (a): the Hertzsprung-Russell diagram. The green region indicates the phase where the super-Eddington wind operates during the first He-shell flash, while the magenta region corresponds to the optically thick wind phase that dominates the subsequent evolution. The red triangle marks the onset of off-center carbon ignition, and the blue filled circle indicates the final evolutionary state. Panel (b): the evolution of the WD radius. The horizontal dashed-dotted lines denote the adopted limits for the optically thick wind prescription. The upper limit (${R}_{\rm {max}}$, red dashed-dotted line) corresponds to ten times the initial WD radius, above which no mass is accreted onto the WD, while the lower limit (${R}_{\rm {min}}$, blue dashed-dotted line) corresponds to twice the initial WD radius, below which all transferred material is accreted. Panel (c): the mass coordinate of the temperature maximum (${M}_{\rm r}({T}_{\rm {max}})$) together with the corresponding temperature ${\log}({T}_{\rm {max}}/{\rm K})$. The red triangle marks the onset of carbon ignition, while the inward movement of ${M}_{\rm r}({T}_{\rm {max}})$ traces the propagation of the off-center carbon flame.}
  \end{center}
\end{figure*}

\begin{figure*}
\begin{center}
\epsfig{file=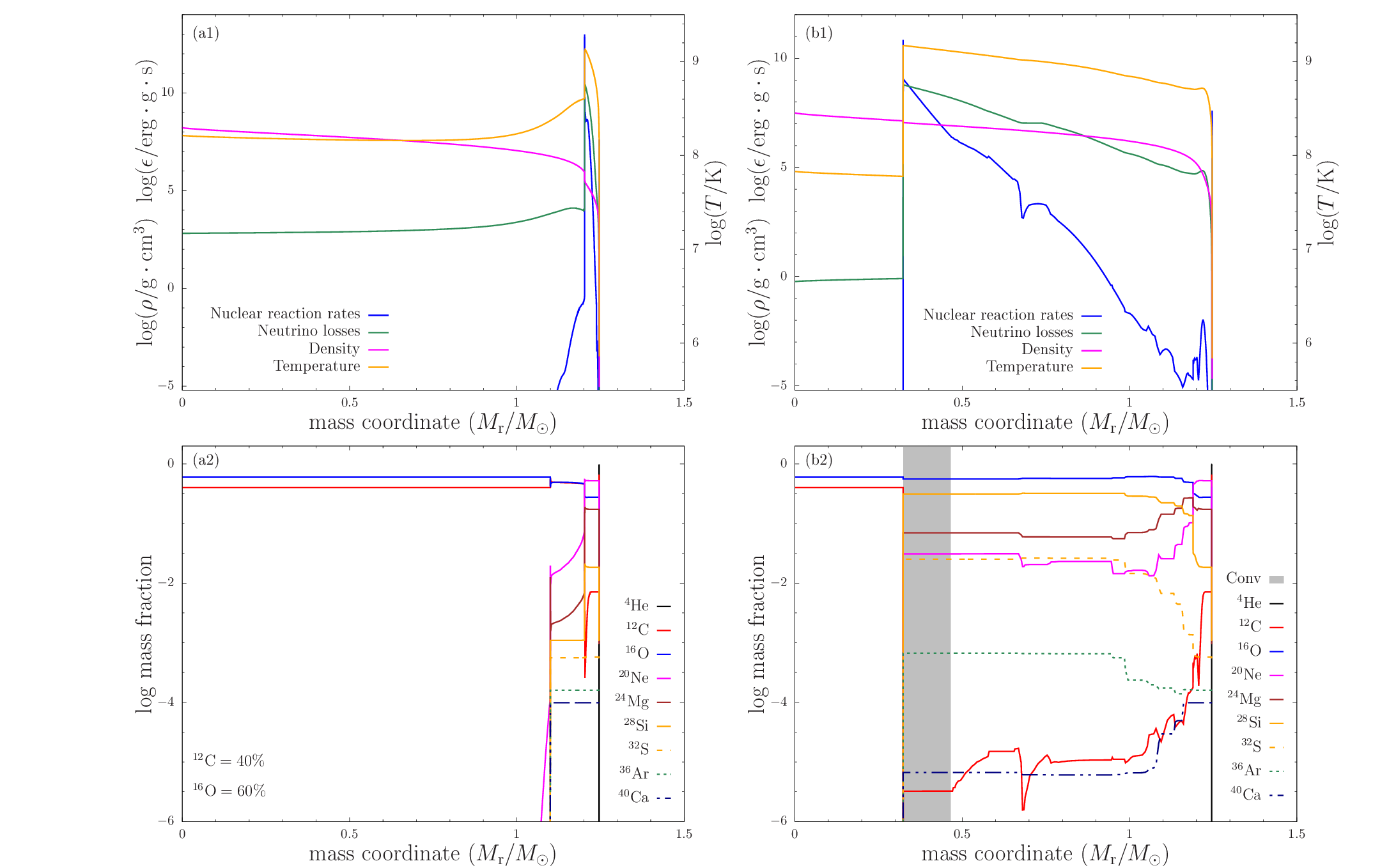,angle=0,width=17.5cm}
 \caption{Internal structure and abundance profiles of the accreting $1.1{M}_{\odot}$ CO WD with an initial composition of $^{\rm {12}}{\rm C}=40\%$ and $^{\rm {16}}{\rm O}=60\%$ at two representative evolutionary stages. Panels (a1) and (a2): the profiles at the onset of off-center carbon ignition. Panel (a1) presents the temperature, density, nuclear energy generation rate, and neutrino cooling rate. Panel (a2) shows the corresponding elemental abundance distribution. Panels (b1) and (b2): similar to panels (a1) and (a2), but for the same quantities at a later stage when the inwardly propagating carbon flame has reached ${M}_{\rm r}\approx0.3{M}_{\odot}$. The grey shaded region in panel (b2) indicates the convective zone associated with carbon burning.}
  \end{center}
\end{figure*}

\begin{figure*}
\begin{center}
\epsfig{file=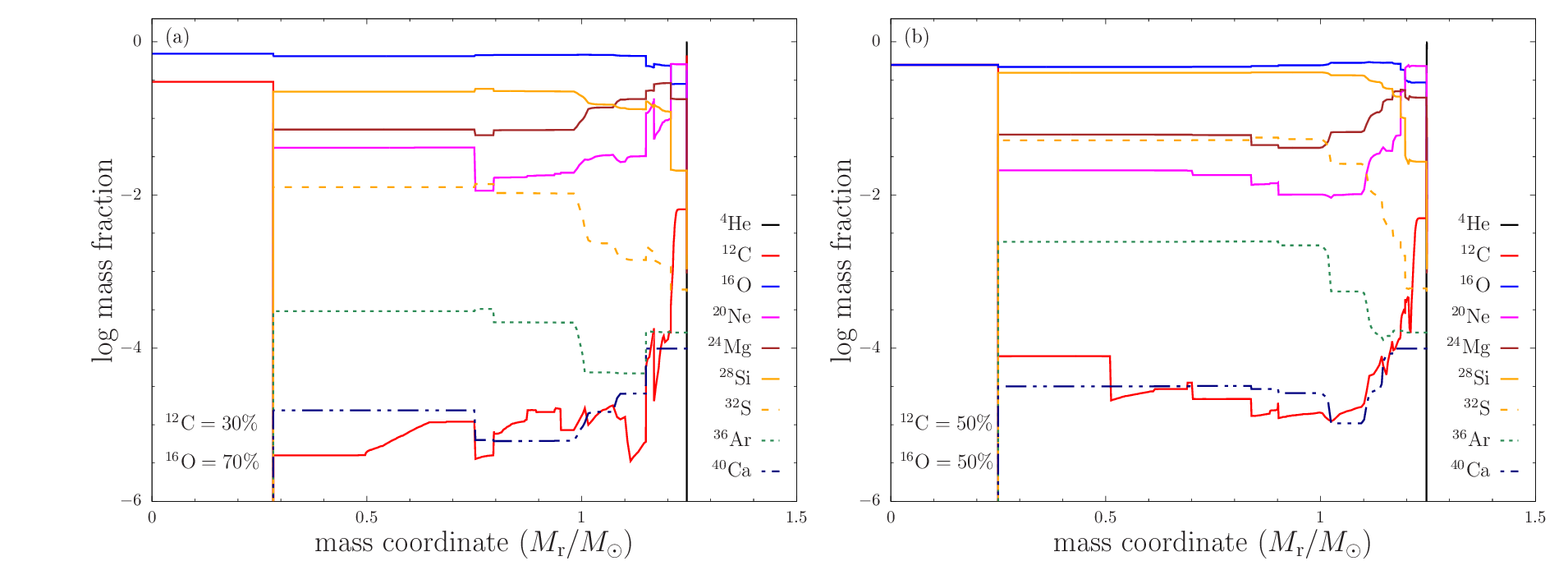,angle=0,width=17.5cm}
 \caption{Elemental abundance profiles of the accreting CO WDs with different initial compositions at the stage when the carbon flame reaches ${M}_{\rm r}\approx0.3{M}_{\odot}$. Panel (a): the model with $^{\rm {12}}{\rm C}=30\%$ and $^{\rm {16}}{\rm O}=70\%$. Panel (b): the model with $^{\rm {12}}{\rm C}=50\%$ and $^{\rm {16}}{\rm O}=50\%$.}
  \end{center}
\end{figure*}

\begin{table*}[htbp]
\centering
\begin{tabular}{|c|c|c|c|c|c|c|c|}
\hline
\multicolumn{2}{|c|}{Observation} & \multicolumn{6}{c|}{Accreting WD models} \\
\hline
 \cline{1-2}\cline{3-8}
\multicolumn{2}{|c|}{SN 2021yfj} & 
\multicolumn{2}{c|}{\begin{tabular}{c}$^{\rm {12}}{\rm C}=30\%$\\$^{\rm {16}}{\rm O}=70\%$\end{tabular}} & \multicolumn{2}{c|}{\begin{tabular}{c}$^{\rm {12}}{\rm C}=40\%$\\$^{\rm {16}}{\rm O}=60\%$\end{tabular}} & \multicolumn{2}{c|}{\begin{tabular}{c}$^{\rm {12}}{\rm C}=50\%$\\$^{\rm {16}}{\rm O}=50\%$\end{tabular}} \\
\hline
Element & Mass fraction & Element & Mass fraction & Element & Mass fraction & Element & Mass fraction \\
$^{\rm 4}{\rm {He}}$ & $-$ & $^{\rm 4}{\rm {He}}$ & $0.41\%$ & $^{\rm 4}{\rm {He}}$ & $0.34\%$ & $^{\rm 4}{\rm {He}}$ & $0.37\%$ \\
$^{\rm 12}{\rm {C}}$ & $-$ & $^{\rm 12}{\rm {C}}$ & $0.097\%$ & $^{\rm 12}{\rm {C}}$ & $0.11\%$ & $^{\rm 12}{\rm {C}}$ & $0.086\%$ \\
$^{\rm 16}{\rm {O}}$ & $78.6\%$ & $^{\rm 16}{\rm {O}}$ & $59\%$ & $^{\rm 16}{\rm {O}}$ & $53\%$ & $^{\rm 16}{\rm {O}}$ & $48\%$ \\
$^{\rm 20}{\rm {Ne}}$ & $10\%$ & $^{\rm 20}{\rm {Ne}}$ & $10\%$ & $^{\rm 20}{\rm {Ne}}$ & $12\%$ & $^{\rm 20}{\rm {Ne}}$ & $11\%$ \\
$^{\rm 24}{\rm {Mg}}$ & $1\%$ & $^{\rm 24}{\rm {Mg}}$ & $17\%$ & $^{\rm 24}{\rm {Mg}}$ & $13\%$ & $^{\rm 24}{\rm {Mg}}$ & $11\%$ \\
$^{\rm 28}{\rm {Si}}$ & $5\%$ & $^{\rm 28}{\rm {Si}}$ & $14\%$ & $^{\rm 28}{\rm {Si}}$ & $21\%$ & $^{\rm 28}{\rm {Si}}$ & $28\%$ \\
$^{\rm 32}{\rm {S}}$ & $3\%$ & $^{\rm 32}{\rm {S}}$ & $0.31\%$ & $^{\rm 32}{\rm {S}}$ & $0.97\%$ & $^{\rm 32}{\rm {S}}$ & $2.2\%$ \\
$^{\rm 36}{\rm {Ar}}$ & $1\%$ & $^{\rm 36}{\rm {Ar}}$ & $0.011\%$ & $^{\rm 36}{\rm {Ar}}$ & $0.024\%$ & $^{\rm 36}{\rm {Ar}}$ & $0.074\%$ \\
$^{\rm 40}{\rm {Ca}}$ & $0.1\%$ & $^{\rm 40}{\rm {Ca}}$ & $0.0043\%$ & $^{\rm 40}{\rm {Ca}}$ & $0.0042\%$ & $^{\rm 40}{\rm {Ca}}$ & $0.0046\%$ \\
\hline
\end{tabular}
\caption{\label{tab:1} Elemental abundance for the CSM of SN 2021yfj inferred from the spectral modeling (first two columns; from \citealt{2025Natur.644..634S}) and the average elemental abundance in the outer $0.3{M}_{\odot}$ of our three accreting CO WD models (columns $3-8$).} 
\end{table*}

\begin{table*}[htbp]
\centering
\begin{tabular}{|c|c|c|c|}
\hline
\multicolumn{4}{|c|}{\begin{tabular}{c}{Accreting WD models}\\$^{\rm {12}}{\rm C}=40\%$\\$^{\rm {16}}{\rm O}=60\%$\end{tabular}} \\
\hline
\multicolumn{4}{|c|}{Nuclear reaction network} \\
\hline
\multicolumn{2}{|c|}{approx21.net} &
\multicolumn{2}{c|}{57iso.net} \\
\hline
Element & Mass fraction & Element & Mass fraction \\
$^{\rm 4}{\rm {He}}$ & $0.34\%$ & $^{\rm 4}{\rm {He}}$ & $0.28\%$ \\
$^{\rm 12}{\rm {C}}$ & $0.11\%$ & $^{\rm 12}{\rm {C}}$ & $0.088\%$ \\
$^{\rm 16}{\rm {O}}$ & $53\%$ & $^{\rm 16}{\rm {O}}$ & $53.5\%$ \\
$^{\rm 20}{\rm {Ne}}$ & $12\%$ & $^{\rm 20}{\rm {Ne}}$ & $10.6\%$ \\
$^{\rm {23}}{\rm {Na}}$ & $-$ & $^{\rm {23}}{\rm {Na}}$ & $0.13\%$ \\
$^{\rm 24}{\rm {Mg}}$ & $13\%$ & $^{\rm 24}{\rm {Mg}}$ & $11.2\%$ \\
$^{\rm 25}{\rm {Mg}}$ & $-$ & $^{\rm 25}{\rm {Mg}}$ & $1.0\%$ \\
$^{\rm 27}{\rm {Al}}$ & $-$ & $^{\rm 27}{\rm {Al}}$ & $0.85\%$ \\
$^{\rm 28}{\rm {Si}}$ & $21\%$ & $^{\rm 28}{\rm {Si}}$ & $22.3\%$ \\
$^{\rm 31}{\rm {P}}$ & $-$ & $^{\rm 31}{\rm {P}}$ & $0.14\%$ \\
$^{\rm 32}{\rm {S}}$ & $0.97\%$ & $^{\rm 32}{\rm {S}}$ & $0.92\%$ \\
$^{\rm 36}{\rm {Ar}}$ & $0.024\%$ & $^{\rm 36}{\rm {Ar}}$ & $0.018\%$ \\
$^{\rm 40}{\rm {Ca}}$ & $0.0042\%$ & $^{\rm 40}{\rm {Ca}}$ & $0.086\%$ \\
\hline
\end{tabular}
\caption{\label{tab:2} Similar to Table\,1, but showing the average elemental abundance in the outer $0.3{M}_{\odot}$ of the fiducial model (columns $1-2$) and of the same model recalculated using a 57-isotope nuclear reaction network (columns $3-4$).} 
\end{table*}

\section{Discussion}

There remain significant uncertainties in the double WD merger scenario for SN 2021yfj. As proposed by \cite{2026arXiv260307064M}, the observed light curve can be reproduced by the interaction between $\sim0.3{M}_{\odot}$ of ejecta and a dense CSM of comparable mass located at a distance of $\sim{10}^{15}\,{\rm {cm}}$, where the CSM is assumed to originate from tidal stripping of the Si-rich WD during the merger. However, current 3D simulations of double WD mergers generally predict much lower amounts of unbound material. For example, by using SPH simulations, \cite{2013ApJ...772....1R} found that only $\lesssim5\times{10}^{-3}{M}_{\odot}$ of material escapes through the L2 point. Similarly, \cite{2014MNRAS.438...14D} showed that the tidal tails typically contain at most a few percent of the total system mass. These results suggest that producing a massive CSM of $\sim0.3{M}_{\odot}$ through tidal disruption alone may be challenging. Observationally, however, there are indications that double-degenerate systems may be capable of producing more extended and massive CSM. For example, some Ca-strong transients (e.g., SN 2019ehk and SN 2021gno), which are possibly related to double WD mergers, show signatures of extended CSM (e.g., \citealt{2022ApJ...932...58J}). More recently, SN 2020aeuh has been interpreted as a Type Ia supernova interacting with a massive CO-rich CSM (e.g., \citealt{2025A&A...704A.135T}). Modeling of its light curve suggests a CSM mass of $\sim1.2{M}_{\odot}$ located at a radius of $\sim3-9\times{10}^{15}\,{\rm {cm}}$, which has been proposed to originate from a double WD merger. Although the formation mechanism of such a massive and extended CSM is still unclear, this event indicates that double-degenerate systems may produce substantially larger amounts of CSM than predicted by current simulations. Therefore, while the requirement of a $\sim0.3{M}_{\odot}$ Si-rich CSM in SN 2021yfj remains difficult to reconcile with existing merger simulations, it may not be entirely ruled out, and further multidimensional studies are needed to clarify this issue.

If the CSM originates from tidal stripping of the Si-rich WD, its composition can be approximated by the average abundances in the outer $\sim0.3{M}_{\odot}$ of the WD. In this scenario, the CSM composition is set by the nucleosynthesis during off-center carbon burning. The properties of carbon burning are sensitive to the mass-accretion rate. A lower accretion rate leads to a slower buildup of the CO layer, allowing more efficient neutrino cooling. This results in a higher density at ignition, which in turn produces a hotter carbon flame and enhances the synthesis of Si-group elements (e.g., \citealt{2019MNRAS.486.2977W}; \citealt{2020MNRAS.495.1445W}). In \cite{2026arXiv260307064M}, the evolution of the accreting CO WD was modeled using a constant mass-accretion rate of $5.0\times{10}^{-6}\,{M}_{\odot}\,{\rm {yr}^{-1}}$ together with a super-Eddington wind prescription. In their model, off-center carbon ignition occurs at a WD mass of $\sim1.261{M}_{\odot}$, corresponding to an average mass-growth rate of $\sim3.6\times{10}^{-6}\,{M}_{\odot}\,{\rm {yr}^{-1}}$, which is slightly lower than that in our simulations ($\sim3.84\times{10}^{-6}\,{M}_{\odot}\,{\rm {yr}^{-1}}$). As a result, their models predict a higher Si abundance (up to $\sim20\%$) in the outer layers of the WD. In contrast, the abundances inferred for the CSM of SN 2021yfj indicate relatively higher O and S but lower Si. This discrepancy may suggest that the progenitor system experienced a different accretion history or had a lower initial C/O ratio. In particular, a combination of lower accretion rates and reduced carbon abundance may help to reconcile the model predictions with the observed abundance pattern, although a more systematic exploration of parameter space is required to confirm this possibility.

In the double WD merger scenario for SN 2021yfj, the observed light curve can be reproduced if approximately $\sim0.3{M}_{\odot}$ of material is ejected through a thermonuclear explosion triggered during the merger process. However, whether the merger process can produce a thermonuclear explosion and unbind a sufficient amount of mass remain uncertain. \cite{2018ApJ...869..140K} performed 3D simulations of a $1.1{M}_{\odot}$ CO WD merging with a $1.2{M}_{\odot}$ ONe WD, and found that the system undergoes a failed detonation, ejecting only a small amount of material ($\sim0.08{M}_{\odot}$) while leaving behind a super-Chandrasekhar ONe WD remnant ($\sim2.2{M}_{\odot}$). This result suggests that thermonuclear burning can be triggered during the merger process, but may be insufficient to produce a supernova with large ejecta mass. The situation may be further complicated in the case considered in \cite{2026arXiv260307064M}. In such scenario, the merging system involves a Si-rich WD formed through off-center carbon burning. As shown by \cite{2016ApJ...832...71L}, the carbon flame is expected to propagate to the center, leaving behind a WD with little residual carbon. Since carbon is the primary fuel for thermonuclear explosions in degenerate conditions, the lack of carbon may significantly affect both the ignition conditions and the explosion strength. Therefore, whether mergers involving such Si-rich WD can produce a successful explosion remains highly uncertain.

If double WD mergers can produce SN 2021yfj-like events, their expected birthrate needs to be assessed. Previous studies have shown that the onset and outcome of off-center carbon burning depend sensitively on the detailed accretion history, which in turn is governed by binary parameters. In addition, in the double WD merger scenario, the Si-rich WD is required to have a mass of at least $\sim1.25{M}_{\odot}$, while its companion ONe WD must be more massive. These constraints significantly restrict the allowed binary parameter space. As a result, the birthrate of SN 2021yfj-like events from this channel is likely to be low. A more detailed investigation of the relevant parameter space and the corresponding birthrate will be presented in a forthcoming study (Liu et al., in preparation).

\section{Summary}

In this work, we simulated the evolution of a $1.1{M}_{\odot}$ CO WD accreting He-rich material at a rate consistent with binary evolution calculations. Under the optically thick wind scheme, off-center carbon ignition occurs when the WD mass reaches $\sim1.246{M}_{\odot}$. The resulting carbon flame attains sufficiently high temperatures to synthesize Si-group elements, producing significant amounts of Ne, Mg, and Si, along with smaller fractions of S and Ar. We further show that the nucleosynthetic yields depend on the initial C/O ratio. A higher initial carbon abundance enhances the production of Si, S, and Ar, while reducing the abundances of O and Mg, whereas Ne and Ca are only weakly affected. Comparing our results with the inferred CSM composition of SN 2021yfj, we find that the outer $0.3{M}_{\odot}$ of the accreting WD generally exhibits overabundances of Mg and Si, and lower abundances of O, S and Ar. Although the predicted abundance does not fully match the observations, our results support that double WD mergers can produce Si-rich WDs and thus remain a viable progenitor channel for SN 2021yfj-like events.

\begin{acknowledgments}

This study is supported by the CAS Project for Young Scientists in Basic Research (YSBR-148), the National Natural Science Foundation of China (Nos 12473032, 12225304, 12288102), the National Key R\&D Program of China (No. 2021YFA1600404), the Yunnan Revitalization Talent Support Program (Young Talent Project, Yunling Scholar Project), the Yunnan Science and Technology Program (Nos 202501AS070005, 202605AS350010, 202601BC070011), the Yunnan Fundamental Research Project (No. 202501AW070001), and the International Centre of Supernovae (ICESUN), Yunnan Key Laboratory of Supernova Research (No. 202505AV340004).

\end{acknowledgments}

\bibliography{citation}{}
\bibliographystyle{aasjournal}

\end{document}